\documentclass[epj,10pt]{svjour}
\usepackage{amsfonts}
\usepackage{color}
\usepackage{eepic}
\usepackage{epic}
\usepackage{ulem}
\usepackage{graphicx}
\usepackage{texdraw}
\usepackage{epsfig}

\usepackage{epsfig}
\usepackage{color}
\begin{document}
\title{Quasi-particle model for lattice QCD: quark-gluon plasma
 in  heavy ion collisions}
\author{Vinod Chandra\inst{1} \thanks{E-mail:\email{ vinodc@iitk.ac.in}},
 V. Ravishankar\inst{1}\inst{2} 
\thanks{E-mail:\email{ vravi@iitk.ac.in}}}
\institute{\inst{1} Department of Physics, Indian Institute of Technology 
Kanpur, UP, India, 208 016\\
\inst{2} Raman Research Institute, C V Raman Avenue, Sadashivanagar, 
Bangalore, 560 080, India}
\date{\today}

\abstract{We propose a quasi-particle model to describe the 
lattice QCD equation of state for pure SU(3) gauge theory in its deconfined state, for $T \ge 1.5T_c$.
The method involves mapping the interaction part of the 
equation of state to an effective fugacity of otherwise
non-interacting quasi-gluons. We find that this mapping is exact. Using the quasi-gluon distribution function, we determine the 
energy density and  the modified dispersion relation for the single particle 
energy, in which the trace anomaly is manifest. As an application, we first determine the 
Debye mass, and then the important transport parameters, {\it viz}, the shear viscosity, $\eta$ and the shear viscosity to entropy density ratio,
$\eta/{\mathcal S}$. We find that  both $\eta$ and $\eta/{\mathcal S}$ are sensitive to the
interactions, and that  the interactions significantly lower both $\eta$
and $\eta/\mathcal S$.}

\maketitle

Quasigluon; Effective fugacity; Debye mass;\\
Shear viscosity; Gluon quenching parameter

\section{Introduction}
The physics of the non-perturbative domain of QCD, unlike the perturbative domain,
is less understood. The physics of confinement and 
quark-hadron transition require a deep understanding of this domain of QCD and is an
area of intense research.
The best known way to address the non-perturbative QCD  is 
the lattice gauge theory\cite{wilson}. One of the important goals 
in lattice QCD is the determination of equation of state(EOS) for strongly interacting matter.
The knowledge of EOS provides a platform
to study many interesting physical phenomena; in particular, at high temperatures, this provides the most realistic EOS for the hot and dense matter(QGP) created in heavy ion collision experiments.

An interesting question that arises is whether the lattice EOS(LEOS) results can be understood
in terms of quasi-particles which are either free, or at most weakly interacting. A positive answer  to this problem would open the doors for developing appropriate effective theories
which can capture the highly non-trivial results of LEOS with a simpler physical picture.
In developing such a picture, an endeavor of this kind 
  may not be expected to yield satisfactory results, if n\"aive parametrizations in terms of quantities such as the effective mass are employed. Rather, they have to be
more in the spirit of the Fermi liquid picture of Landau \cite{landau} where the energy is
a complicated functional of the number density. We undertake a similar exercise here, for pure gauge theory, and show that such a description can indeed be obtained in terms of excitations which may be looked upon as quasi-gluons -- with an effective fugacity which captures all the interaction effects. We find that our agreement with the lattice results is not merely qualitative;
its deviation is less than one part in a million. The method employed here uses and elaborates upon
the model introduced earlier by us \cite{chandra1,chandra2,chandra3} for studying hot pQCD EOS. 

As an application of this effective description, we investigate LEOS predictions for the viscosity $\eta$, and the viscosity to entropy ratio $\eta/{\mathcal S}$. These transport parameters are central to the understanding of the properties of QGP which is produced in heavy ion collisions. Indeed, recent experimental observations\cite{star} from RHIC strongly suggest 
that QGP created at RHIC behaves like a near perfect fluid, having a
very low viscosity to entropy ratio, $\eta/{\mathcal S} \ge 1/4\pi$ \cite{star,vis1,shur,son}.
This implies  that at temperatures close to $T_c$, the quark matter in the QGP  phase is strongly interacting, and is perhaps  in the non-perturbative domain of 
QCD.  These findings are in accordance with the  lattice studies which  predict that the 
hot QCD equation of state is approximately $10\%$ away from its ideal counterpart even at 
$T=4T_c$\cite{boyd,karsch,gavai,cheng}. It should, therefore, be natural to employ LEOS to determine the transport parameters. However, theoretical studies \cite{arnold,asakawa1} seek  by  treating the equilibrium state to be that of an ideal gas of quarks and gluons.
Such an assumption does not seem to be justified in view  of the lattice results. Consequently, the determination of $\eta/{\mathcal S}$ requires a revisit where the non-ideal nature of the EOS is explicitly incorporated. Further, since its determination
is 
best undertaken in terms of a transport equation\cite{arnold,asakawa1,xu}, the quasi-gluon picture lends itself naturally to undertake that exercise.

In addition to studying $\eta/{\mathcal S}$, we  employ the quasi-particle picture to extract
the Debye mass, via the transport equation. This exercise allows us to determine the value
of the phenomenological coupling constant that occurs in the Yang-Mills and the Vlasov terms
in the transport equation. As an indication of the robustness of the model, we are able to get a complete
agreement between the lattice and the quasi-particle results. We make a few remarks in passing on the implications to heavy quark dissociation in QGP.

Yet another quantity of interest is the bulk viscosity, which survives provided that the trace anomaly is non-vanishing. We note that since the quasi-particle representation is exact, it automatically reproduces the trace anomaly. It is therefore possible to determine, in principle, the bulk viscosity as well by using the tarnsport approach. It would be of great interest to compare the results so obtained with those obtained in Refs.\cite{kharzeev1,kharzeev2}. This study will be undertaken separately.

The paper is organized as follows. In Section 2, we introduce
the quasi-particle model,  and extract the 
equilibrium distribution function from the pure lattice gauge theory EOS, 
and discuss physical meaning of the effective fugacity. 
In Section 3, we  study the temperature dependence of the number density of the quasi-gluons.
By plugging in this expression in the non-abelian Vlasov equation, we determine the Debye mass
and also the value of the phenomenological coupling constant. Using this we further  estimate the dissociation temperatures for heavy quark systems.
In Section 4,  we determine the temperature dependence of gluon quenching parameter, $\hat{q}$. We further determine  the shear viscosity $\eta$ , and the ratio $\eta/{\mathcal S}$. We do find a very small value for $\eta/{\mathcal S}$, as the experiments suggest.
In fact, we find that it can violate the AdS/CFT  (KKS) bound $\frac{1}{4\pi}$\cite{son}. In section 5, we present the conclusions and future prospects.  The mathematical details of the determination of $\eta$ for LEOS has been shown in the Appendix. 

\section{Quasi-particle model for pure gauge theory EOS}
\subsection{The effective fugacity}
We now propose a quasi-gluon description of LEOS at high temperatures. As mentioned in the introduction, our approach is in the spirit of Landau's Fermi liquid theory\cite{landau}. 
The quasi-particle description has been introduced by us in \cite{chandra1,chandra2}. It has been further used in \cite{chandra3} to determine $\eta$ and $\eta/{\mathcal S}$. Yet, the salient features of the model were not fully covered in the earlier papers, which we do so here in the following.

The basic idea is to describe the quasi-particles-- the quasigluons-- by a Bose Einstein distribution (see Eq.(\ref{eq1})). As mentioned,  the analogy with the ordinary bosons is formal. This is so since, as in the Fermi liquid theory, the single particle energy
of the quasi-gluons (which define the distribution) is itself a functional of the number density. This functional, which establishes the collective nature of the response, is to be determined by employing the lattice equation of state. 

We implement the description by writing the distribution function for otherwise {\it free} quasi-gluons in terms of an  effective fugacity $z_g$.  The effective fugacity contains all the interaction effects, and contributes to the energy of the gluons in a non-trivial manner. The
success of the prescription is established {\it a posteriori}. We obtain an exact mapping, and as we show below, the notion of the temperature dependent effective mass which has been employed earlier, is realized only in some limiting situations. We caution that the fugacity which we introduce is merely to establish the relation between the number density and the
energy of the quasi-gluons. In short, our problem tantamounts to determining $z_g$ self consistently
from LEOS.

With the   grand canonical distribution function in mind,  we write the
equilibrium distribution for the quasi-gluons as
\begin{equation}
\label{eq1}
f^g_{eq}=\frac{z_g\exp(-\beta\epsilon_p)}{(1-z_g\exp(-\beta\epsilon_p))}
\end{equation}
where the quantity $(\epsilon_p=p)$ would be the energy of the gluons in the absence of interactions.
The expression for $E_p$, the energy of the quasi-gluons will be determined below.
It may be noted that Eq(\ref{eq1}) is not the same as the distribution function which would follow from a n\"aive adaptation of the Fermi liquid theory.

 On the other hand, the  grand canonical partion function in terms of the effective fugacity may be 
written as follows,
\begin{equation}
\label{eq2}
\ln(Z)=-\nu_g\frac{V}{(2\pi)^3}\int d^3p \ln(1-z_g\exp(-\beta p)), 
\end{equation}
where $\nu_g=2(N_c^2-1)$ is the number of  degrees of freedom for gluons, and $V$ is the volume. 
As a strategy to determine $z_g$,   
we Taylor expand the partition function around $z_g=1$ (ideal gluon gas),
and determine the fugacity by comparing it with LEOS, order by order.
We find that it is sufficient to expand upto $O({\delta z}^2)$, where $\delta z=z_g-1$.
We obtain
\begin{equation}
\label{eq3}
\ln(Z)=\ln(Z_I)+A_1 \frac{V\nu_g}{2\pi^2\beta^4}\delta z+A_2 \frac{V\nu_g}{2\pi^2\beta^4}(\delta z)^2 + O[(\delta z)^3],
\end{equation}
where  $\ln(Z_I)= V\frac{8\pi^2}{45\beta^4}$ is the ideal partition function. The coefficients  $A_1$ and $A_2$ are given in terms of the following integrals,
\begin{eqnarray}
\label{eq4}
A_1&=&\int_0^\infty du u^2 \frac{\exp(-u)}{(1-\exp(-u))}\equiv 2\zeta[3]\nonumber\\
A_2&=&\int_0^\infty du u^2 \frac{\exp(-2u)}{(1-\exp(-u))^2} \equiv \frac{1}{3}[\frac{\pi^2}{3}-6\zeta[3]].
\end{eqnarray}

\begin{figure}[htb]
\begin{center}
\vspace*{-70mm}
\hspace*{-42mm}
\psfig{figure=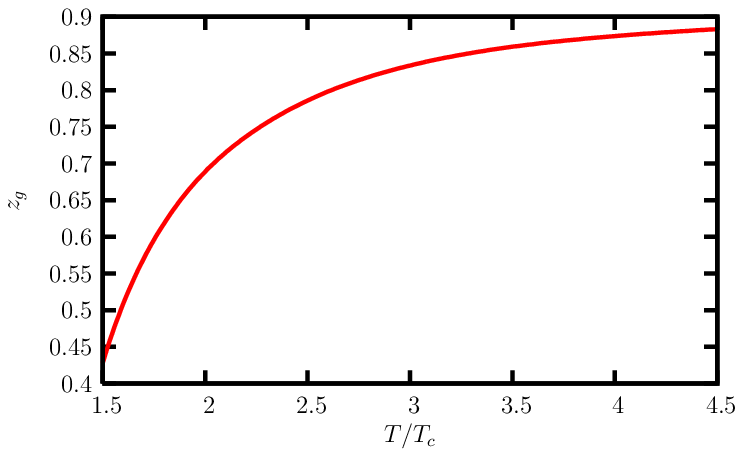,width=150mm}
\vspace*{-80mm}
\caption{\label{fig1} (Color online) Behavior of $z_g$ as a function of $T/T_c$.}
\end{center}
\end{figure}

It is straight forward to obtain the expression for the pressure from the partition function  Eq.(\ref{eq3}) via 
\begin{equation}
\label{eq5}
P_g\beta V=\ln(Z).
\end{equation}
Rewriting the lattice expression for the pressure as
 $P_L=P^I_g+\Delta P_g$
(where $P^I_g=\frac{8 T^4\pi^2}{45}$ is the ideal part of the pressure and $\Delta P_g$ accounts for the non-ideal part of the pressure) and match with RHS of  Eq.(\ref{eq5}), we obtain the following quadratic equation for $\delta z$,
\begin{equation}
\label{eq6}
A_2 {\delta z}^2+A_1\delta z-\frac{2\pi^2 \Delta P_g}{\nu_g}=0.
\end{equation}
This equation posses two solutions for $\delta z$:
\begin{equation}
\label{eqz}
\delta z=\frac{-A_1}{2 A_2}\bigg (1\mp\sqrt{1+ \frac {8 A_2\pi^2 \Delta P_g}{\nu_g A^2_1}}\bigg).
\end{equation}
Of the two roots written above, only the first root is physically acceptable. This follows from the requirement 
that $\vert\delta z\vert <1$ and the facts that the discriminant in Eq.(\ref{eqz}) is positive and that the ratio
$A_1/{2 A_2}\approx -0.92$.  This choice also has the correct limit when $\Delta P_g=0$ ($z_g=1$).

We have plotted the effective fugacity ($z_g$) from  Eq.(\ref{eqz}) as a  function of temperature in Fig.1. From Fig.1, it is easy to see that $z_g$ attains its 
ideal value only asymptotically and $0<z_g<1$. 
More importantly, it is clear from Fig.2 that the quasi-gluon description of LEOS is exact when $T \ge 1.5T_c$. The deviations are negligible, being of  $O(10^{-6})$ . This agreement assures the reliability of our results for observables such as viscosity with the quasi-gluon picture. 

\subsubsection{Physical significance of the effective fugacity}
The physical significance of the effective fugacity introduced in the present paper is different from that of the effective mass employed in \cite{pesh1,pesh2,allton,pm}. We show below that the effective mass description of \cite{pesh1,pesh2,allton} emerges from our more general framework only as a  limiting case. In fact, $z_g$  regulates the number density as a function of temperature,  apart from contributing to the dispersion relations for the quasi-gluons. We analyze the latter feature first.

\begin{figure}[htb]
\begin{center}
\vspace*{-70mm}
\hspace*{-42mm}
\psfig{figure=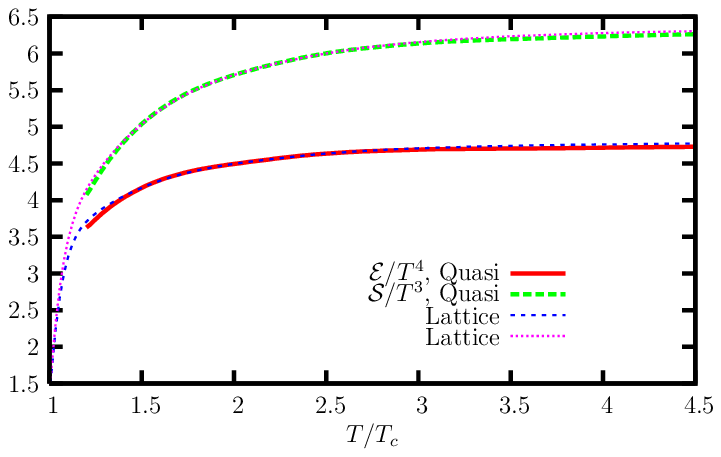,width=150mm}
\vspace*{-80mm}
\caption{\label{fig2}(Color online) Energy density and entropy density as a function of temperature determined from the quasi-particle model. The corresponding lattice results are also shown. The quasi-particle results show almost perfect quantitative agreement with the lattice results.}
\end{center}
\end{figure}
\subsubsection{The dispersion relation}
Notwithstanding appearances, the energy of the quasi-gluons is not merely given by the relation
$\epsilon_p=p$.  Rather, it should be determined from the fundamental thermodynamic relation between the energy density 
and the partition function

\begin{equation}
{\mathcal E}_g=-\frac{1}{V}\partial_\beta Ln(Z_g).
\end{equation}
Substituting  Eq.(\ref{eq3}) for the partition function $Z_g$, we obtain the following interesting expression,
\begin{equation}
\label{edn}
{\mathcal E}_g=\frac{\nu_g}{8\pi^3}\int d^3p [p+T^2\partial_T \ln(z_g)] f^{g}_{eq}.
\end{equation}

The modified dispersion relation for a quasi-gluon reads,
\begin{equation}
\label{des}
E_p=p+T^2\partial_T \ln(z_g).
\end{equation}

Notably, we see that the dispersion relation  has picked  up an additional contribution, 
$T^2\partial_T \ln(z_g)$, which is purely temperature dependent. Note that the usual fugacity terms for free bosons do not contribute to the dispersion relation.
The additional  term is crucial since it owes its emergence to the nonvanishing trace anomaly in LEOS.
Therefore, this additional (purely temperature dependent) scale, which  gives the non-zero conformal 
measure,  is responsible for  the bulk viscosity. Interestingly, the presence 
of this scale does not change the velocity of the gluons, since $v_g=\partial_p E_p$.

We now study the situations under  which the effective mass prescription would be viable.
 To that end, we  cast  Eq.(\ref{des}) in the form 
\begin{equation}
 (p+T^2\partial_T \ln(z_g))^2 \equiv (p^2+m^2),
\end{equation}
which leads to the identification
\begin{equation}
\label{ems}
 m^2(T)=2p T^2\partial_T \ln(z_g) +(T^2\partial_T \ln(z_g))^2.
\end{equation}
 The first term in the expression 
for $m^2(T)$ is linear in the momentum apart from being temperature dependent, while the
second  is purely temperature dependent. Thus, if  $z_g$ is to be realized in terms of 
an effective mass, the mass would have to be momentum  dependent. However, in the low momentum limit( ultra soft quasi-gluons), the first term becomes subdominant  {\it wrt} the second term. In this particular limit, the effective fugacity can be interpreted as a purely temperature dependent effective mass. The condition translates to $p \ll T^2\partial_T \ln(z_g)$. From this we see that RHS of Eq.(\ref{ems}) $\rightarrow 0$ as $T \rightarrow \infty$.

Integrating the RHS of Eq.(\ref{edn}), we obtain the energy density as,
\begin{equation}
 {\mathcal E}_g=3P_g + \Delta_g,
\end{equation}
where$\Delta_g= T^2\partial_T \ln(z_g){\cal N}_g$ is the trace anomaly and ${\cal N}_g$,
 is the quasi-gluon number density,
\begin{equation}
 {\cal N}_g=\frac{\nu_g}{8\pi^3}\int d^3p f^g_{eq}.
\end{equation}
We shall study ${\cal N}_g$ in detail in the next section.

Before we end this section, we note that effective fugacity descriptions have been earlier
employed in condensed matter systems in the last decade. We summarize these works briefly.  To study the nature of Bose-Einstein (BE) condensation transition
in interacting Bose gases, a parametric EOS in terms of the effective fugacity has been 
proposed   by Li {\it et al}\cite{li}, which provides  a 
 scheme for exploring the quantum-statistical nature of the BEC transition with interacting gases. Effective fugacity has been used for a 
unitary fermion gas  by Chen {\it et al}\cite{chen} for studying 
 thermodynamics with  non-Gaussian correlations. Purely as technical tool to distinguish the
populations in the condensate state from the others, effective fugacity has been employed by Haugerud {\it et al,}\cite{haugd}  for a BE system of non-interacting bosons in a  harmonic trap. A similar approach has been employed in Refs.\cite{haugset,noh,kirsten} for studying BEC  with interacting bosons. None of them  employs the effective dispersion relation which we obtain naturally in this work.

\subsubsection{Entropy Density}
The entropy density as a function of temperature can be obtained from Eq.(\ref{eq2}), by employing
${\mathcal S}=\frac{1}{V}\partial_T \ln(Z)$. After some straightforward  
manipulation, we get
\begin{eqnarray}
\label{eqen}
{\mathcal S}= 4\frac{P_g}{T}+\frac{\Delta_g}{T}. 
\end{eqnarray}
The first term in the above equation is due to the unmodified dispersion relation, while the second  term  is nothing but the trace anomaly contribution to the 
entropy density.  

The behavior of the  energy density and that of the entropy density are shown in Fig. 2. As expected, they match with the lattice results, displaying   the viability of the quasi-particle model. It will be seen in section 4 that the  temperature dependence of ${\mathcal S}$  will make a substantial contribution to the temperature dependence of the  ratio, $\eta/\mathcal{S}$ for QGP. 
\section{The effective number density and Debye mass}
\subsection{The number density}
We turn our attention to the number density of the quasi-gluons, which need not be the same as that of the interacting gluons.  It is given by
\begin{equation}
{\cal N}_g=\frac{\nu_g}{8\pi^3}\int d^3p f^g_{eq}(p,z_g). 
\end{equation}
 Using the isotropy of the distribution function and performing the momentum integral one obtains,
\begin{equation}
\label{eqn}
{\cal N}_g=\frac{\nu_g}{\pi^2\beta^3}{\mathcal PolyLog}[3,z_g].
\end{equation}
Its ideal counter part reads ($z_g=1$),
\begin{equation}
\label{eq0}
{\cal N_I}=\frac{\nu_g}{\pi^2\beta^3}\zeta[3],
\end{equation}
where the function ${\mathcal PolyLog}[n,z_g]\equiv\sum_{k=0}^{\infty} z_g^k/k^n$.
In Fig.3, we plot the ratio of  the  number density of the quasi-gluons relative to that of ideal gluons,  $R_{{\cal N}}={\cal N}_g/{\cal N_I}$,
as a function of temperature. The ratio is always less than unity and approaches the ideal limit asymptotically. 

\begin{figure}[htb]

\begin{center}
\vspace*{-70mm}
\hspace*{-42mm}
\psfig{figure=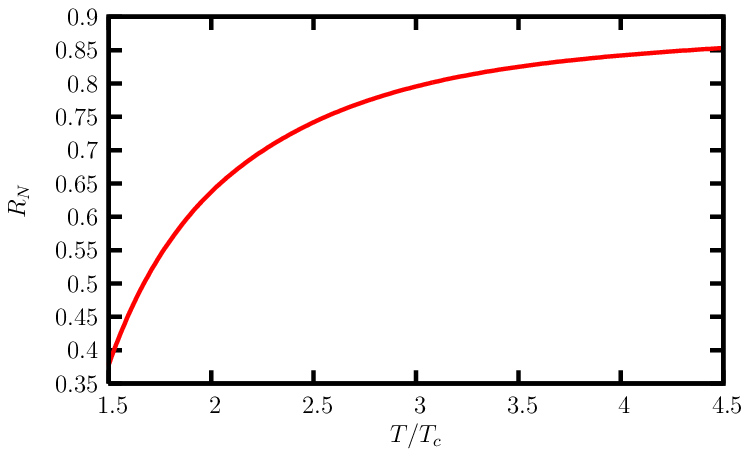,width=150mm}
\vspace*{-80mm}
\caption{\label{fig3} (Color online) Behavior of $R_N$ as a function of $T/T_c$.}
\end{center}
\end{figure}

\subsection{The Debye mass}
The Debye mass is independently determined by lattice computations, and as such, there is no need to address it again within our model. Yet, we may ask if the knowledge of $M_D$ can throw light on the effective coupling constant $g^{\prime}$  which occurs in the transport equation. A determination of the effective coupling constant is warranted since it contributes to the transport and the thermodynamic properties of QGP. In particular, the viscosity -- which we are interested in this paper -- depends on $g^{\prime}$ (see Eq.(\ref{eq26})). 

To that end, we  employ the equilibrium distribution obtained in the previous section to
and write the permittivity at zero frequency in the form $\tilde{\epsilon}(\omega,k) = 1 +M_D^2/k^2$, in terms of the Debye mass $M_D$\cite{manual,akranjan,chandra2}, which is given by
\begin{equation}
\label{eqm}
M^2_D=-2N_c(g^\prime)^2\int d^3p \partial_{\epsilon_p}f^g_{eq}(p,z_g), 
\end{equation}
which, on an explicit evaluation acquires the form
\begin{equation}
\label{eqm}
M^2_D=(g^{\prime})^2 \beta^{-2} \frac{2N_c}{\pi^2} {\mathcal PolyLog}[2,z_g].
\end{equation} 

We now match the  Debye mass  in Eq.(\ref{eqm}) with the
lattice parametrized expression for the Debye mass, $M_D^{L}=(1.40)g(T) T$ employed in\cite{mocsy}. This allows us to identify  $g^\prime$ to be,
\begin{equation}
g^\prime =\frac{1.40 g(T)\pi}{\sqrt{6 {\mathcal PolyLog}[2,z_g]}}.
\end{equation}
Incidentally, the plasma frequency 
$\omega_p =M_D/\sqrt{3}$. 

\subsubsection{Dissociation temperatures for quarkonia}
We make a brief digression to estimate the dissociation temperature for heavy quarkonia, as predicted by LEOS.
Recall that a quarkonium state is stable against  strong decay
if the over all mass of the pair of quarks remains below the open charm and beauty thresholds. The large masses of 
the charm quark($m_c\sim 1.5 Gev$) and the bottom quark($m_b=4.5 GeV$) allow a study of their spectroscopy, based on the non-relativistic(NR) potential theory\cite{pot}.  One favorite choice of the potential in the confined phase is the Cornell potential,
\begin{equation}
V(r)=\sigma r-\frac{\alpha}{r}
\end{equation}
 where $\sigma\sim 0.2 GeV^2$ and $\alpha\sim \pi/12$ are the phenomenological parameters. Employing this 
form of the potential in the NR Schr\"odinger equation\cite{hsatz}
 leads to the  values of the radii($r_{q\bar{q}}$) of various quarkonia states as listed in Table 1.
For the complete list of energy and mass of various charmonium and botomonium states, we refer the reader to Ref.\cite{hsatz}. Note that these numbers obtained from a NR theory give a good account of quarkonium spectroscopy (the masses are determined with an less than 1\% error for all spin averaged states).

In the QGP phase, due to the  
the screening of chromo-electric field, the quarkonioum bound states survive up 
to a temperature. One simple way to determine the temperature at which a particular state dissociates is: whenever $1/M_D \le r_{q\bar{q}}$, where $r_{q\bar{q}}$ is
the {\it rms}  radius of the state, the particular 
state will not survive in the medium. 
 The equality yields the dissociation temperature $T_d$, which we display in table 2,
by employing the Cornell potential. These estimates are smaller than the other estimates
for $T_d$ \cite{chandra2,hsatz,alberico,sdata,gert,umeda}, and somewhat close to the results
obtained in Ref.\cite{mocsy,chandra4}. But this can perhaps not be taken too seriously since the
criterion for determining $T_d$ requires refinement.

\begin{table}
\caption{\label{table1} Radius for various quarkonia 
states (in unit of fm) taken from Ref.\cite{hsatz}.}
\centering
\begin{tabular}{|l|l|l|l|l|l|l|l|l|l|l|}
\hline
  $q\bar{q}$ state &$J/\Psi$ &$\chi_c$&$\Psi^\prime$&$\Upsilon$&$\chi_b$&$\Upsilon^\prime$&$\chi^\prime_b$&$\Upsilon^{\prime\prime}$\\
\hline
 $r_{q\bar{q}}$ in fm &0.25&0.36&0.45&0.14&0.22&0.28&0.34&0.33\\
\hline
\end{tabular}
\end{table}

\begin{table}
\label{table1}
\caption{\label{table2} Dissociation temperature($T_d$) for various quarkonia 
states (in unit of $T_c$). Note that $T_c$ is taken to be 0.27 GeV\cite{zantow}. We employ  2-loop expression for the 
QCD running coupling constant at finite temperature\cite{schoder} }
\centering
\begin{tabular}{|l|l|l|l|l|l|l|l|l|l|l|}
\hline
 $q\bar{q}$ state &$J/\Psi$ &$\chi_c$&$\Psi^\prime$&$\Upsilon$&$\chi_b$&$\Upsilon^\prime$&$\chi^\prime_b$&$\Upsilon^{\prime\prime}$\\
\hline
$T_d/T_c$ &1.24&1.00&1.00&2.56&1.47&1.07&1.00&1.00\\
\hline
\end{tabular}
\end{table}

\section{The shear viscosity}
We now consider the important physical quantity,
the shear viscosity $\eta$ and its ratio to the entropy density, $\eta/{\mathcal S}$. 
Determination of $\eta$ requires a knowledge of the collisional properties of the medium
when it is perturbed away from equilibrium.
Of the two methods that  determine the 
transport parameters, {\it viz.} the Kubo formula, and the semi-classical  transport theory,
we adopt the latter one in this paper, and follow the approach 
of Asakawa {\it et al.} \cite{asakawa1}. 

The shear viscosity  has two contributions\cite{asakawa1}, (i) from the Vlasov term which captures the long range component of the interactions, and (ii) the collision term which models the short
range component of the interaction. The net viscosity is given by
 $1/\eta=1/\eta_A +1/\eta_C$, where the first term gets 
its contributions from the diffusive Vlasov term  and the second term gets contribution 
from the collision term.  
Asakawa, Bass and M\"uller \cite{asakawa1}
have argued that the diffusive Vlasov contributions to the shear viscosity dominates 
in the weak coupling limit. We restrict our study to determine $\eta_A$
and $\eta_A/{\mathcal S}$ here. We shall  drop the subscript $A$ hence forth.

In their work, Asakawa, Bass and M\"uller \cite{asakawa1}  have considered the 
Vlasov term for an ensemble of turbulent color fields, but assume that the equilibrium configuration is that of an ideal gas of gluons. Such an assumption is clearly not admissible
while employing LEOS. In an earlier paper, we  have generalized their work
to  a perturbatively interacting QGP \cite{chandra3}. It was found that the inclusion of interactions significantly decreases $\eta$ and
$\eta/{\mathcal S}$. LEOS is expected to cause similar significant changes, which we estimate now.

\begin{figure}[htb]
\begin{center}
\vspace*{-75mm}
\hspace*{-42mm}
\psfig{figure=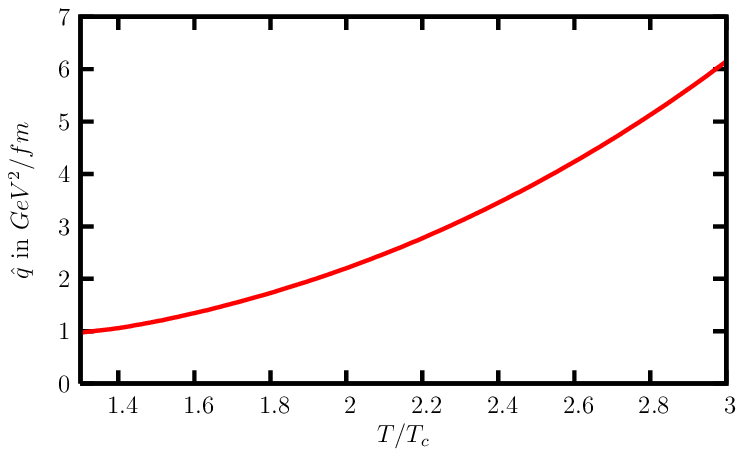,width=150mm}
\vspace*{-82mm}
\caption{\label{fig4}  (Color online)  Gluon quenching parameter ${\hat q}$ as a function of $T/T_c$.
}
\end{center}
\end{figure}

\begin{figure}[htb]
\begin{center}
\vspace*{-75mm}
\hspace*{-42mm}
\psfig{figure=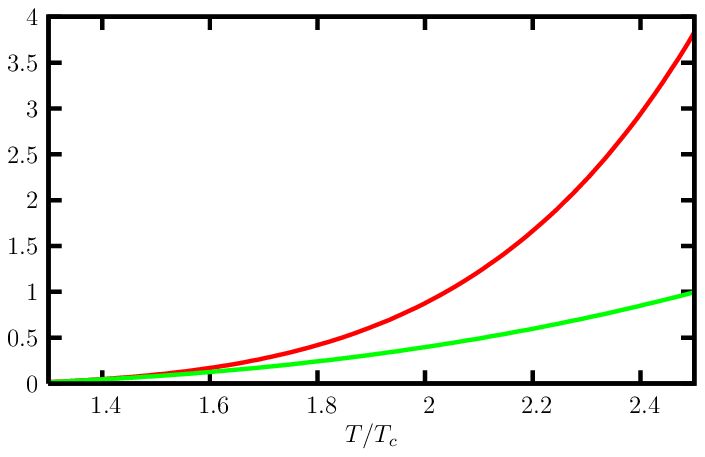,width=150mm}
\vspace*{-82mm}
\caption{\label{fig5} (Color online)  The quantity $3\pi^2\eta/{32(N_c^2-1)T^3_c}$  as a function of $T/T_c$ is shown.
The upper curve corresponds to constant $\hat{q}=1 GeV^2/fm$ and the lower curve is obtained by employing the 
temperature dependence of $\hat{q}$.}
\end{center}
\end{figure}

The method of obtaining $\eta$ has been
described at length in \cite{landau1,asakawa1,chandra3}. Using the same method,
we may write,
\begin{equation}
 \eta=\frac{-\beta}{15}\int \frac{d^3p}{8 \pi^3} \frac{p^4}{E^2_p} \bar{\Delta}( p) \partial_{E_P} f_{eq}(p),
\end{equation}
where $\bar{\Delta}( p)$ parametrizes the anisotropy in the distribution(for details see Ref.\cite{chandra3}).
  $\bar{\Delta}( p)$ can be determined by the variational procedure from the 
linearized transport equation\cite{landau1,asakawa1} with a Vlasov term and
a collision term  computed by Arnold {\it et al}\cite{arnold}. It is important to 
note that the work of Asakawa {\it et al}\cite{asakawa1}
is the generalization of the work of based on the earlier work of Dupree\cite{dupree} 
for the non-abelian plasmas.

The form of $\bar{\Delta}(p)$ employing ideal EOS has been determined in \cite{asakawa1} in the case of a 
purely chromo-magnetic plasma, and they obtain
\begin{equation}
\label{eq24}
 \bar{\Delta}( p)=\frac{ (N^2_c-1) E^2_p T}{3 C_2 (g^\prime)^2 <B^2> \tau^{mag}_m},
\end{equation}
where  $g^\prime$ is the phenomenological coupling and $\tau^{mag}_m$ is
the magnetic relaxation time. We demonstrate that  expression for $\bar{\Delta}( p)$ in the present case is formally the same as the one given above.  This follows from the fact $z_g$, which captures all the interaction effects in $f_{eq}$ is independent of momentum and  is purely temperature dependent. Recall that accordingly, the expression for the particle energy gets modified to $E_p=p+T^2\partial_T [\ln(z_g)]$ (see Eq.(\ref{des})), and that the energy density is related to the pressure via ${\cal E}=3 P + \Delta$. Keeping these in mind, the same procedure as in \cite{asakawa1} may be followed which yields Eq.(\ref{eq24}). The details are given in the appendix.

The expression for $\bar{\Delta}( p)$ taking $\tau_m$ along the light cone\cite{abhijit} would then be:
\begin{equation}
\bar{\Delta}( p)=\frac{ (N^2_c-1) E^2_p T}{3 C_2 (g^\prime)^2 <E^2+B^2> \tau_m}.
\end{equation}
Here, the lightcone frame is introduced only to relate the denominator of the above equation with the gluon quenching parameter.
It must be borne in mind that $ <E^2+B^2>$  is essentially the energy density which must be
determined by taking only the contributions  from the soft modes.
 Thus, 
\begin{equation}
\label{eq26}
\eta= \frac{(N^2_c-1)\beta}{15\pi^2 C_2 (g^\prime)^2<E^2+B^2> \tau_m} \int_0^\infty dp \ p^{6} f_{eq}(1+f_{eq}).
\end{equation}
Employing the  distribution function Eq.(\ref{eq1}), extracted from LEOS,
one obtains the following expression for the shear viscosity,
\begin{equation}
\eta=\frac{(N^2_c-1)^2}{15\pi^2 N_c (g^\prime)^2 <E^2+B^2> \tau_m}\int_0^\infty p^6 \frac{z_g\exp(-\beta p)}{(1-z_g\exp(-\beta p))^2},
\end{equation}
which after performing the  momentum integral becomes,
\begin{equation}
\eta=\frac{16 (N^2_c-1)^2}{\pi^2 N_c (g^\prime)^2 <E^2+B^2> \tau_m} T^6 {\mathcal PolyLog}[6,z_g].
\end{equation}
Note that the shear viscosity for ideal gluons  is given by,
\begin{equation}
\eta^I= \frac{16 \zeta(6) (N^2_c-1)^2}{\pi^2 N_c} \frac{T^6}{(g^\prime)^2 <E^2+B^2> \tau_m}.
\end{equation}
In the above expression, 
$<E^2 +B^2>$
gets contributions from the soft modes, and is hence not the standard energy density.
The Debye mass is a convenient parameter to demarcate  the soft and the hard modes, whence we
perform the momentum integration in Eq.(9) only up to $M_D$. Denoting the resulting energy density by ${\cal E}^{S}$, we obtain

\begin{eqnarray}
\label{ens}
 \frac{\cal E^{S}}{T^4}&=&\frac{\nu_g}{2\pi^2}\bigg(\sum_{l=1}^\infty \frac{z_g^{l}}{l^4} \gamma[4,1.4 g(T) l]
+ T\partial_T[\ln(z_g)]\nonumber\\&&\times \sum_{l=1}^\infty \frac{z_g^{l}}{l^3} \gamma[3,1.4 g(T) l]\bigg),
\end{eqnarray}
where $\gamma[n,x]$ is the lower incomplete gamma function: 
$\gamma[n,x]=(n-1)!(1-\exp(-x)\sum^{n-1}_{k=0} x^k/k!)$.

 Since the parameter $\tau_m$ is unknown, one approach 
 is to relate  $<E^2 +B^2>\tau_m$ to the gluon quenching parameter ${\hat q}$\cite{abhijit} as given by
\begin{equation}
\hat{q}= \frac{16\pi \alpha_s N_c}{3(N^2_c-1)} {\cal E^{S}} \tau_m. 
\end{equation}
in writing which we have employed the relation $\frac{<E^2 + B^2>}{2}={\cal E^{S}}$,
which follows from LEOS. In that case, the expression for shear viscosity reads,
\begin{equation}
\label{eqv}
\eta=\frac{32 (N^2_c-1) T^6 {\mathcal PolyLog}[6,z_g]}{3\pi^2 \hat{q}}
\end{equation}
The expression for the ratio $\eta/{\mathcal S}$ can be obtained by combing Eq.(\ref{eqv})
and Eq.(\ref{eqen}).

Clearly, what can be determined in this approach unambiguously is the ratio
\begin{equation}
\label{eqhq}
 \frac{{\hat q}}{\tau_m}= \frac{16 \pi \alpha_s N_c}{3(N_c^2-1)} \cal E^{S}.
\end{equation}

Thus, we see that in the approach taken above, the problem of determining $\eta$ reduces to a determination of either the gluon quenching parameter $\hat{q}$, or of $\tau_m$. There are several attempts to determine $\hat{q}$. The approach based on the twist expansion \cite{quench1} predicts a value $\hat{q} \sim 1 -2 GeV^2/fm$ at a temperature $T_0 \sim (337\pm 10) MeV$. The estimation based on the eikonal approximation \cite{quench2} predicts a much larger range of values, between $10 -30 GeV^2/fm$. Note that the above estimates, which are fitted from the data, are by no means precise, and are available only at one particular value of  temperature. The value of $\eta$ also inherits the same uncertainty.

While it is not easy to eliminate the uncertainty in $\hat{q}$ in the above mentioned analyses, we show that it is possible to determine its temperature dependence, given its value at some temperature, say at $T_0$. We do so by considering the parameter $\tau_m$ instead. For the plasma under consideration,
an  intuitively appealing  way is to 
relate $\tau_m$  to the plasma frequency $\omega_p$,  as $\tau_m=C \omega^{-1}_p$, where $C$ is the  proportionality constant.

\begin{figure}[htb]
\begin{center}
\vspace*{-75mm}
\hspace*{-42mm}
\psfig{figure=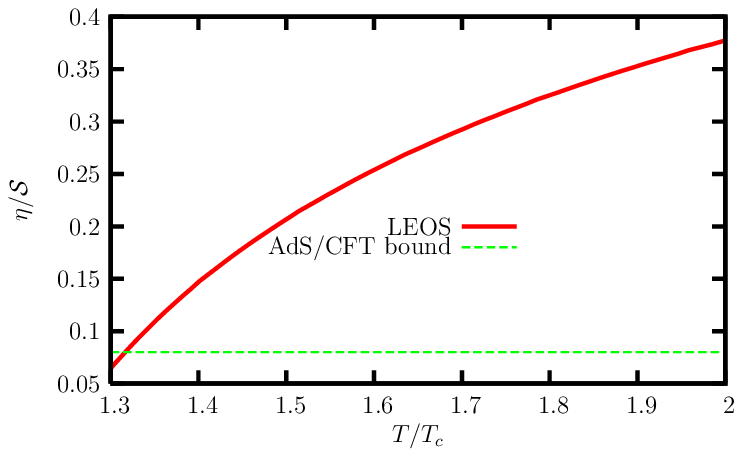,width=150mm}
\vspace*{-82mm}
\caption{\label{fig6} (Color online) Viscosity to entropy density ratio ($\frac{\eta}{{\mathcal S}}$) as a function of $T/T_c$.  Note that, we have chosen $T_c=.27 GeV$ \cite{zantow}.}
\end{center}
\end{figure}
The plasma frequency, $\omega_p$ for LEOS can be determined by employing the quasi-particle model
in the expression for the chromo-electric susceptibility, in the limit $ k \rightarrow 0$:
 $ \tilde{\epsilon}(\omega,0)=1-\frac{\omega^2_p}{\omega^2}$. 
It is easy to check that $\omega_p=M_D/\sqrt{3}$. At this point 
we employ the expression for the Debye mass determined in the previous section. 
It is important to note that $M_D$, and hence $\omega_p$, are sensitive to the interactions, which makes our ansatz plausible.
On the other hand,
we fix $C$  by using the value of $\hat{q}$ at $T_0 $. Since its  value has been estimated to be in the range $1 -2Gev^2/fm$  \cite{quench1},  we get
 $C \approx 1.07 -2.14$. This data point completely fixes $\hat{q}$ as a function of temperature. We have not employed the other set of values since they are not easy to accommodate within the perturbative frame work which we have employed here.
 We have shown $\hat{q}$ as a function of temperature in Fig. 4. It is clear that $\hat{q}$ has a strong dependence on temperature which cannot be ignored in the determination of $\eta$.

We have plotted the shear viscosity, $\eta$ for LEOS as a function of temperature in Fig. 5, with the choice $C=1.07$.
For comparison we have also shown the values of $\eta$ when $\hat{q}(T)$ is assumed to be a constant. The strong dependence of $\hat{q}$ on the temperature is clearly reflected in the viscosity, with its value getting lowered substantially around $T=2.5T_c$. 
\begin{figure}[htb]
\begin{center}
\vspace*{-75mm}
\hspace*{-42mm}
\psfig{figure=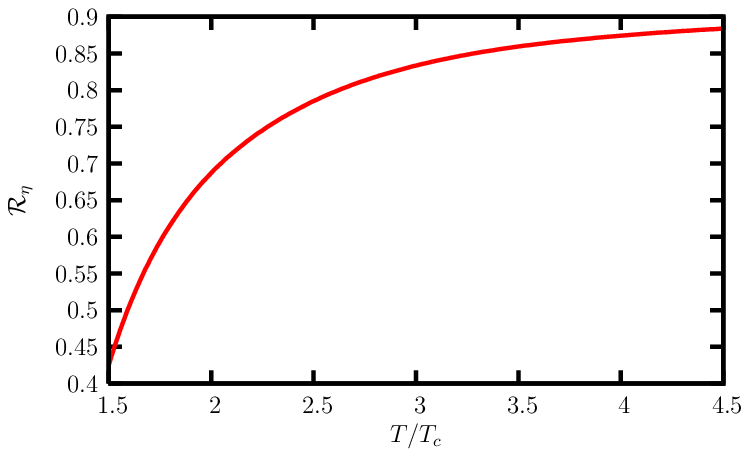,width=150mm}
\vspace*{-82mm}
\caption{\label{fig7} (Color online) ${\mathcal R}_\eta$ as a function of $T/T_c$.}
\end{center}
\end{figure}
We have shown the behavior of $\eta/{\mathcal S}$ as a function of temperature in Fig. 6.
It appears that the ratio may  violate the   the AdS/CFT bound, $1/{4\pi}$, marginally for $T \le 1.5 T_c$.  A larger violation of the bound is possible at higher temperatures, if one employs  eikonal based estimates for $\hat{q}$.

Let us consider the ratio ${\mathcal R}_\eta=\eta/\eta^I$ to see how $\eta$ for  LEOS deviates from 
its ideal counterpart. This ratio is model independent to the extent that it does not depend on $\hat{q}$. In making this statement it is understood that as a phenomenological parameter, $\hat{q}$ is not sensitive to the EOS employed \cite{quench1,quench2}. The behavior of  ${\mathcal R}_\eta$ as a function of $T/T_c$ is shown in Fig.7, from which 
it is clear that the inclusion of interactions significantly decreases the shear viscosity.  
As expected, the ratio ${\mathcal R}_\eta$ asymptotically approaches unity. 
Therefore, the shear viscosity serves as a 
good diagnostic to distinguish the EOS at RHIC.

To see the extent to which the interactions effect the $\eta/{\mathcal S}$,  we consider the ratio ${\mathcal R}_{\eta/{\mathcal S}}=\frac{\eta/{\mathcal S}}{\eta^I/{\mathcal S}^I}$.
The behavior of ${\mathcal R}_{\eta/{\mathcal S}}$ as a function of temperature is shown in Fig.8. From Fig.8 it is easy to see that interactions coming from LEOS decrease the ratio $\eta/{\mathcal S}$ by $\approx$ 35\% near $1.5 T_c$ and $\approx 5\%$ near $3 T_c$. It approaches the corresponding ideal value only asymptotically.
This crucial observation reinforces the necessity of employing realistic equations of state, in particular LEOS for determining the transport properties of the plasma.
 Our findings lead to an interesting conclusion that both $\eta$ and the ratio, $\eta/\mathcal S$ are good diagnostics as far as the effects of interactions are concerned.

\begin{figure}[htb]
\begin{center}
\vspace*{-70mm}
\hspace*{-42mm}
\psfig{figure=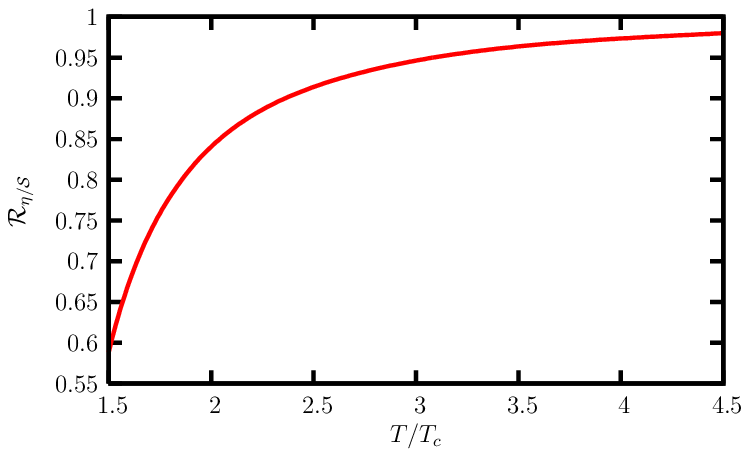,width=150mm}
\vspace*{-80mm}
\caption{\label{fig8} (Color online)  Viscosity to entropy density ratio relative to its ideal counterpart as a function of $T/T_c$. Note that this plot is model independent since the ratio, ${\mathcal R}_{\eta/{\mathcal S}}$ is interdependent of ${\hat q}$}
\end{center}
\end{figure}

We further note that we recover the expression for the ratio $\eta/{\mathcal S}$ 
obtained by Majumder {\it et al }\cite{abhijit}S if we take the limit  $z_g\rightarrow 1$, 
as a special case.  Our results for $\eta/{\mathcal S}$ are at a variance  with the predictions of \cite{meyer,nakamura,adare,gavin,chandra3,visco,vis1,lacey,grner,xu,xu1}.

\section{Conclusion and Outlook}
In conclusion, we have developed a quasi-particle model in the 
spirit of Landau's Fermi liquid theory  to extract
the distribution function for gluons from pure gauge theory 
equation of state. We find that the description is exact.
We show that all the interaction effects
can be captured in the effective fugacity for gluons.
We have determined the new dispersion relation for 
quasi-gluons which brings out the effect of trace anomaly and also the collective
nature of these excitations.
We have determined the temperature dependence of the Debye mass which can be 
exactly matched with the lattice parametrized Debye mass by defining the effective 
gluon charge in terms of the QCD running coupling constant.
Employing the quasi-particle model, we have determined $\eta$ and the ratio $\eta/{\mathcal S}$. 
In doing this, we have determined $\hat{q}$ as function of temperature for LEOS.
We have also determined the temperature dependence of gluon quenching parameter.
We find that both  $\eta$ and ${\eta/\mathcal S}$ for LEOS decrease significantly
as compare to the ideal EOS. We find that there is a possible violation of AdS/CFT bound for 
$\eta/{\mathcal S}$ for lattice equation of state.
It would be of interest to extend this analysis to the full QCD EOS and also to study the bulk viscosity.
Should the quasi-particle model work for full QCD equally well, it opens up interesting possibilities of building effective theories.

\vspace{1.5mm}
\noindent
{\bf Acknowledgment}:
We are thankful to Frithjof Karsch  for providing us with the lattice data which made this analysis possible.
VC acknowledges Saumen Datta for useful  discussions 
and Ashok Garai for help in the numerical part. He acknowledges the Raman Research Institute, Banglore (India) for hospitality where  part of this work was completed, and C.S.I.R, New Delhi  (India) for financial support.

\section{Appendix}
In this appendix, we show how one determines the anisotropy parameter, $\bar{\Delta}(p)$ for LEOS.
We start with the equilibrium distribution function 
$f_{eq}=1/(z^{-1}_g\exp(\beta u.p)-1)$, where $z_g$ is purely temperature dependent, for the quasi-gluons. The action of the drift operator on $f_{eq}$ is given by
\begin{eqnarray}
(v\cdot\partial)f_{eq}&=&-f_{eq}(1+f_{eq})\bigg\lbrace(p-\partial_{\beta} \ln(z_g))v\cdot \partial (\beta)
\nonumber\\&&+\beta (v\cdot \partial)(u\cdot p)\bigg\rbrace,
\end{eqnarray}
 where we recognize that $p-\partial_{\beta} \ln(z_g)\equiv E_p$,  is the modified dispersion relation. In the local rest frame of the fluid, this expression is formally the same 
 as Eq.(6.1) in \cite{asakawa1} where of course $E_p=p$.  Similarly, the expressions for the
 Debye mass and  the continuity equation for the energy momentum tensor ( Eqs.(6.3)-(6.7) in \cite{asakawa1})
  also undergo the same modification  via the new dispersion relation.  

The final expression for the drift term after imposing the energy-momentum conservation is obtained as
\begin{eqnarray}
\label{eqd} 
(v\cdot\partial)f_{eq}(p)&=&f_{eq}(1+f_{eq})\bigg[\frac{p_i p_j}{E_p T} (\nabla u)_{ij}\nonumber\\&&-\frac{m^2_D E^2 \tau^{el}_m E_p}{3T^2 {\partial {\mathcal E}}/{\partial T}}\nonumber\\&&+(\frac{p^2}{3E^2_p}-c^2_s)\frac{E_p}{T}(\nabla\cdot\vec{u})\bigg],\nonumber\\ 
\end{eqnarray}
where $c^2_s$ is the speed of sound. 
The third term in Eq.(\ref{eqd}) will contribute to the bulk viscosity.
To determine the  bulk viscosity for LEOS, we need to include trace part in the ansatz for $f_1(\vec{p},\vec{r})$\cite{asakawa1}.
In this case, the form of the perturbation, $f_1(\vec{p},\vec{r})$ gets modified as,
\begin{equation}
f_1=-\frac{p_ip_j}{E_p T^2}\bigg( (\nabla u)_{i j}\Delta_1(p)+\delta{ij}\frac{1}{3}(\nabla\cdot \vec{u})\Delta_2(p)\bigg).
\end{equation}
The second term in the above expression will generate a term proportional to $\nabla \cdot\vec{u}$ in the force term, and the comaprison of this term with the third term in Eq.(\ref{eqd}) would lead to the expression for $\Delta_2(p)$ and hence the bulk viscosity. Since we are only interested in the shear viscosity here, 
we concentrate on the form of $\Delta_1(p)\equiv\bar{\Delta}(p)$.

On the other hand, the force term will have exactly the same mathematical form as in \cite{asakawa1}(Eq.(6.13)), if we consider only the 
traceless part of velocity gradient in the expression for $f_1(\vec{p},\vec{r})$.
The same mathematical structure of the Force term in this case  follows from the isotropy of $f_{eq}(p)$. 
The force term in the case of a purely chromomagnetic plasma in the present case will be,
\begin{eqnarray}
 \nabla_p\cdot D^{mag}\cdot \nabla_p \bar {f}(p)&=&\frac{3 C_2 \bar{\Delta}(p) B^2 \tau^{mag}_m}{(N^2_c-1)E^3_p T^2}\nonumber\\&&\times f_{eq}(1+f_{eq}) p_i p_j
(\nabla u)_{ij}
\end{eqnarray}

On comparing of the Force term and the first term in the RHS of Eq.(\ref{eqd}), we infer that
the anisotropy parameter is given by
\begin{equation}
\bar{\Delta}(P)=\frac{(N_c^2-1)E^2_p T}{3C_2 g^2 B^2 \tau^{mag}_m}
\end{equation}

\end{document}